# A framework of reusable structures for mobile agent development

Tudor Marian

Computer Science Dept.

Tech. Univ. of Cluj-Napoca

26-28, G. Baritiu St.
400027 Cluj-Napoca

Romania

*tudorm@coned.utcluj.ro*

Bogdan Dumitriu

Computer Science Dept.

Tech. Univ. of Cluj-Napoca

26-28, G. Baritiu St.
400027 Cluj-Napoca

Romania

*bdumitriu@bdumitriu.ro*

Mihaela Dinsoreanu

Computer Science Dept.

Tech. Univ. of Cluj-Napoca

26-28, G. Baritiu St. 400027 Cluj-Napoca

Romania

*Mihaela.Dinsoreanu@cs.utcluj.ro*

Ioan Salomie

Computer Science Dept.

Tech. Univ. of Cluj-Napoca

26-28, G. Baritiu St.
400027 Cluj-Napoca

Romania

*Ioan.Salomie@cs.utcluj.ro*

*Abstract –* **Mobile agents research is clearly aiming towards imposing agent based development as the next generation of tools for writing software. This paper comes with its own contribution to this global goal by introducing a novel unifying framework meant to bring simplicity and interoperability to and among agent platforms as we know them today. In addition to this, we also introduce a set of agent behaviors which, although tailored for and from the area of virtual learning environments, are none the less generic enough to be used for rapid, simple, useful and reliable agent deployment. The paper also presents an illustrative case study brought forward to prove the feasibility of our design.**

## I. INTRODUCTION

Although agent platforms are becoming increasingly widespread and powerful these days, although they are evolving rapidly towards what some call a "second generation" and although they are the centerfold of extensive research all around the world, there still appears to be something which prevents them from being universally adopted as a natural evolution of the object oriented world. In our opinion, there are actually two things, not just one, which can be held responsible for the current situation: lack of simplicity and lack of interoperability.

Let us discuss lack of simplicity first. Most agent platforms nowadays find themselves in one of the two extremes: they either offer enormous flexibility at the cost of usability (the user gets tangled in aspects that are completely irrelevant to her application, thus losing perspective) or they offer extended built-in functionality at the price of interoperability and extensibility (the user has plenty of predefined agent services to choose from and plug into her application, but is confined to the specific agent platform). The JADE [7] and the ADK [8] agent platforms both fall into one of the two above categories. Neither of these extremes can completely satisfy the user's requirements. Ideally, an agent platform should offer the means for the user to easily set up and run her agents (if so required) and also the flexibility of changing virtually everything about an agent if the need arises.

We also mentioned lack of interoperability in the argument above. This is a bit more difficult to see, especially for the untrained eye, due to the standardization efforts such as those of FIPA [6] which allow a certain degree of interoperability between agents built on top of different agent platforms, or due to included semantic support, such as that provided by ontologies. The catch, however, is that compatibility among agents belonging to different frameworks which are FIPA-compliant is virtually restricted to what FIPA defines. This is a serious drawback, if one cares to analyze it, because FIPA defines nothing in terms of components which can be used for rapid agent development and deployment. The implication of this fact is that implementation efforts employed by various agent platform developers are not reusable outside their own platform (at least not without serious adaptations efforts). The ultimate goal should be for the user to be able to simultaneously benefit not only from a single, but from all development work conducted by researchers and implementors all around the globe.

These are the two main deficiencies which we have tried to address with our work. Our idea was to overcome such limitations by defining an abstract, platform independent, framework which would allow the user to easily:

- define her agents in terms of high level abstractions;
- be able to "plug in" various common predefined behaviors;
- switch from one platform to another as the need arises.

Given that the area where we make use of agents falls roughly into the category of computer based education, the idea described before was primarily focused on the development of a framework which would allow us to benefit from all the aforementioned improvements in our line of work. As a consequence, we narrowed our agent directed research down to field of virtual learning environments (VLE). This does not mean, however, that we didn't design our framework to be as general as possible. It merely implies the fact that the behaviors we have identified and introduced into our implementation were drawn mainly from the VLE field. Even so, by reading the paper, it will become obvious that most of behaviors described are equally applicable to a variety of other fields as well.

In short, a virtual learning environment represents a space available on-line where both students and teachers are brought together to interact similarly to the way they do in reality. There are a few things pertaining to a VLE which allowed us to make certain simplifying assumptions in our work. In the first place, a VLE is usually what one calls a "closed environment", meaning it has few (if any) interactions with other systems. This has a direct consequence in that our agents will usually be confined to using "interior" (as opposed to "border") protocols only. Secondly, strong security requirements are normally absent from a VLE, which makes concerns for agent communication security a rather unimportant issue in our case. Finally, the tasks our agents need to perform are generally simple ones and can easily be achieved by the composition of two or more basic behaviors, thus enabling us to disregard some of the more advanced behaviors (which might indeed be appropriate for other environments) from our analysis.

We have based our agent work on that explained in [4]. What we have done was pick up where the authors of [4] left off and change the agents employed in the assessment service (AS) described there in order to make it benefit from the advantages of our agent framework. This was actually a means to prove the feasibility of our research.

In the first part of our paper we introduce and justify the need for simple agent based reusable structures, especially in an environment such as a VLE. We do this by describing our vision of a unifying agent platform framework that puts more (sample) real platforms all under the confines of the same abstraction. We also bring forward a behavior taxonomy which, in our opinion, can successfully be used both in a VLE and in other environments to create significantly useful agents "on the fly". The detailed description of such a behavior (namely, the *Itinerary* behavior) is also addressed in last subsection of this part.

Since we have also made use of our platform in practice as well, the second part of this article is dedicated to a case study which covers the realization of a VLE assessment service based on the agent technology. The study places the service within the virtual university we have created, defines the technical challenges that we have encountered and discusses the solutions we have proposed. We close up with some conclusions and ideas we still intend to pursue in this area.

## II. AGENT BASED REUSABLE STRUCTURES IN VLE

Reusable structures [5] are indeed a very powerful tool aiding the rapid and well-structured development of applications, and it is our belief that agent-based software should benefit of such mechanisms too. We have tried to extract the behavioral patterns that agents are enacting while deployed in their environment, namely the virtual learning environment (VLE).

It is well known that complexity severely compromises the usability of models, therefore reducing the chances of adoption. Indeed, simple things have the reputation of offering a better solution, if applicable, since they are both easier for the user to understand and, therefore, use and for developers to implement. The actual challenge was to find the balance between expressive power and simplicity. This required the analysis of the VLE domain we have previously explored in [3] and [4] in order for us to:

- determine the set of adequate abstractions for our conceptual model
- avoid artificial inclusion of complex abstractions, especially of those that may easily be composed on top of the basic ones, as well as ones that may be needed on rare, possibly particular, occasions

To achieve this goal we strived to find the minimal set of features and abstractions that would be both useful and powerful enough to naturally describe the complex behaviors of coexisting agents working together to perform an assigned task.

### A. Unifying Agent Platform Framework

Recurrently occurring patterns in the use of agent-based development led us to pinpoint certain behaviors encountered on numerous occasions, enforcing the idea of reusable structures when handling this paradigm as well.

These structures were embodied into a comprehensive agent behavioral model shaped on top of a unifying framework. By means of such a framework we managed to make the agent platform transparent to the user and, in the same time, decouple the reusable patterns from the underlying mobile agent platform. It thus becomes clear that the model was structured to be highly independent, encompassing a handful of abstract features that allow it to be equally expressive regardless of the underlying agent support.

Entities common to every agent platform (location, agent, message, behavior, agent identifier along with other relevant ones) provide the context within which we were able to define the reusable patterns. These patterns produce an environment that ultimately separates the behavioral model from the actual skeleton upon which the patters are enacted (i.e. the JADE agent platform) and, as such, once they are created, rewriting them will not be necessary for every new platform. Simply put, one has only to write new adapters if needed, or use the available ones along with the already existing framework items to integrate (coalesce) the component she requires.

Adapters were employed to provide the bridge between the framework and agent platforms. In this way, adding support for a new system only requires the writing of the appropriate wrappers. The following diagrams (Fig. 1 and Fig. 2) illustrate the most relevant adapter hierarchies and their position in the system.

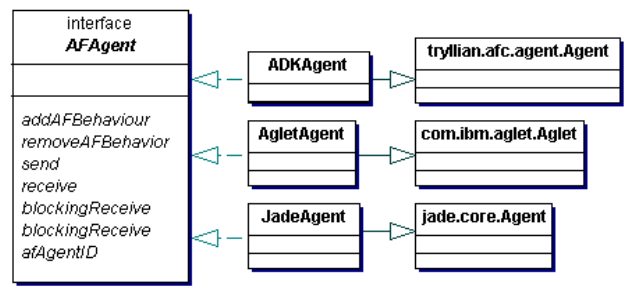

Fig. 1. Agent framework class diagram – agent adapters

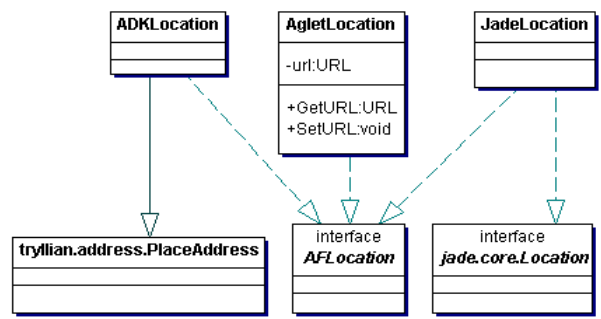

Fig. 2. Agent framework class diagram - location adapters

Other components are adapted similarly, following the general line of the architectural pattern. We have refrained from including more diagrams due to space issues.

The ultimate purpose of the framework was to provide a fast and easy method of deploying agents by connecting together the necessary pieces, in our case behavior patterns. These patterns are to be presented in a more detailed manner during the following section.

### B. Agent Behavior Pattern Taxonomy

The basic patterns we have identified can be grouped under headings such as creation-based (Role Factory, Clone), task-oriented (Observer, Task), communicational (Client-Server, Listener) and mobility-related patterns (Itinerary). Extending this hierarchy should be a fairly easy task, and one that we encourage as well, given that we only provide those concepts related to the VLE context – but their overall level of generalization should not be overlooked when developing a particular solution. The following figure (Fig. 3) shows the hierarchical structure of the behaviors.

*Role Factory*: It allows adding a behavior to an agent dynamically, at runtime, the agent being unaware of the concrete behavior type it will enact. This particular behavior can be used to provide additional behaviors to various agents, behaviors that these agents should be unaware of at the time of their creation. It can be used, for example, in a scenario where a social group of agents is created, performs some sort of initial task and later awaits to be assigned a distinct behavior by this factory.

*Itinerary:* It provides the agent with the ability to perform given tasks at several locations along its route. Basically, it enables the agent to travel from one destination to another, attempting to best meet a preestablished schedule at the same time. For example, this kind of behavior can be used in route discovery scenarios. Although some could argue that it is rather VLE domain specific, we believe that this pattern is useful throughout various scenarios, and our assessment service provides a pertinent example in this direction. It is why it shall be detailed in the later sections.

*Observer:* This pattern allows the given agent to check periodically for the occurrence of various events. Should any such event occur before the next time quanta has elapsed, an event handler will be triggered, performing the required action. It can be used when a number of tasks have to be performed preconditioned by particular events. For example, consider a VLE assessment service and the agent that is given the task of delivering the mandatory tests to recipients at given dates (the midterm date). Such an agent can use this behavior in the following manner: if the time has come (the trigger in this case would be the reaching of a certain date) the action to be taken is the delivery of tests to students (one could use the Itinerary behavior for such a task).

*Listener:* It is a communication pattern - it brings the agent it belongs to in a state of waiting for a message. The message is set up with a type, thus enabling the agent to be awoken only when a message with the expected type arrives (naturally, one can use a generic type to grab every message received). Only then will all the registered event callbacks be fired. Such a behavior can be used in a large number of scenarios, for instance when a message addressed to an agent is supposed to originate a certain action (or sequence of actions). Multiple event listeners can be registered in order to be fired on the arrival of a message; this behavior can be used either cyclically or in a "one shot" manner.

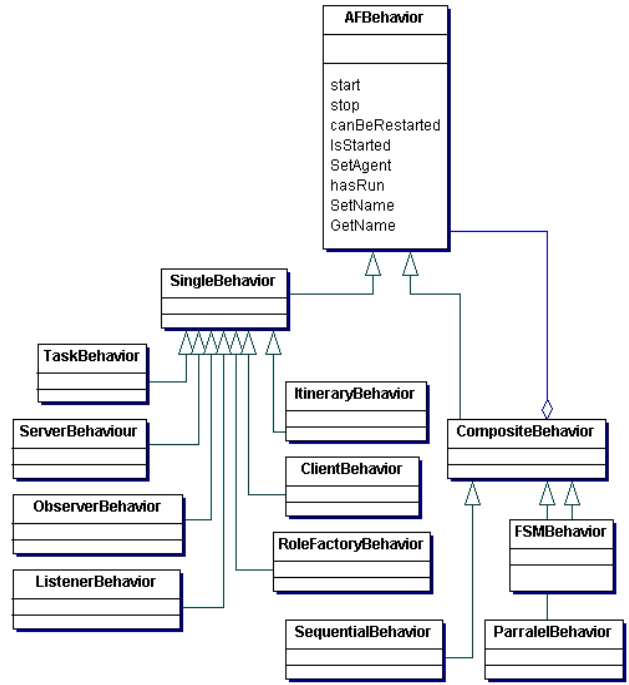

Fig. 3. Agent framework behavior hierarchy

*Client – Server:* The two roles are connected, since both require their counterpart to properly perform its task within the communication protocol. Since the messages are considered to have a high degree of reliability, the protocol requires only one acknowledgment from the server (actually a server worker thread) towards the client. Both the sever and the client take turns in communicating with each other, and the semantic is as follows:

- The *Client*, knowing the location of the server, initiates the communication protocol, by sending a request. This request encapsulates a task the server is required to perform (and a container for the result of the operation). The client blocks awaiting an acknowledgment message, and, if received, it awaits the result of the task performed by the server worker. Timeout is used to detect a possible failure and the agent terminates the behavior (to emphasize the flexibility of our framework let us mention that one could, for example, use this behavior together with the *Observer* to retry connecting after a certain period of time has elapsed). The acknowledgment message is required in this phase because it it the most vulnerable period within the sequence, when the server could not be available due to different adversities.
- The *Server* behavior, when started, awaits requests from its clients. Upon the arrival of a valid message a new server worker thread is spawned (enabling multi threaded server response policy) which takes over all further communication with that particular calling client. Each server worker sends an acknowledge message back to the originator of the message, namely the corresponding client, starts processing the task he was given, again by the client along with the initial message, and sends back the result. This result message is the one that brings the client behavior to its end, thus terminating the current communication session with the server (the thread).

This pattern is susceptible to variations and further adaptations; we tried to reach a fair level of functionality without imposing too many bounds to the model, as that would most likely discourage the user from its employment.

*Task:* This is the standard "one shot" behavior expiring after the assigned job has completed. It is a very general, simple pattern one can use in every situation, and should be considered when extending simple behaviors.

While we have tried to identify the main abstractions one could use, surely there are others we have failed to recognize, or that may become relevant in the future. In addition, since these abstractions are extracted solely from the VLE context, horizontal as well as vertical domains may recognize the need for adding several other relevant components, whether they are more general or more domain-specific. Consequently, the model was devised to be extensible in that it allows the adding of any newly identified property (or behavior for that matter) to the framework with very little effort and no modifications.

Clearly, one can infer the need for composition when handling such basic behaviors in order to increase the complexity and also the power of the model. As a response to this, three such types of behaviors have been introduced: sequential, parallel and Finite State Machine (FSM), and more can be added as the circumstances require it.

- Sequential behavior deploys its elements one at a time, following a predefined, dynamically changeable, order;
- Parallel behavior deploys all its elements on separate "threads". The components are enacted in a parallel manner;
- FSM conforms to a somewhat sophisticated description of state machines (the composite branch can be observed in Fig. 4 as well). States and transitions are the main concepts used when deploying this behavior, while the *Task* pattern is used to model the activity per state.

## C. Itinerary Pattern Description

To better comprehend the inner workings of the behaviors, in the following we shall provide a more thorough description of one such entity, namely the Itinerary Behavior, that was deliberately under-described in the above section to avoid redundancy. It is meant to provide agents with the ability to travel along a path following a time schedule and perform assigned tasks at each location, the locations being part of its initial route description.

Basically, to deploy this pattern one has to initially configure the behavior with a *Route*, several *ObjectiveReachedListener* items, and an *ObjectiveMissedBehavior*. None of these elements can be changed at any time after the behavior has been enacted. The sequence diagram and the state-chart diagram (Fig. 4 and Fig. 5) offer a graphical representation of the entities, relations, flow of messages and main states defining this behavior. Hopefully they will provide enough insightful ancillary information for one to better understand our design approach.

*The Route* is a list of objectives; each objective contains a location that must be reached and a pair of time indicatives, the earliest and the latest possible time of arrival at that given location. The time units are increments to a base value established when the behavior is initially created. This solution was adopted due to the nature of the agent environment's possible distribution (since one agent could be required to travel at different locations, possibly on different time zones to perform its assigned task) and also due to the need to consider the variable travel delay and the delay raised by the deactivation and reactivation of the agent before and after the migration process. Should the agent reach one location on schedule, i.e. at some time between the two above mentioned time indicatives, the *ObjectiveReachedListener* registered listeners are fired. Upon completion the agent migrates to the next location, chosen to best meet the deadline of arrival. If the agent reaches one location early it enters a wait state until enough time elapses, awakes and performs the assignments, then continues along its path. If, on the contrary, one location has not been reached on time (the delays are considered, although ways of properly computing them are still under development) the agent is free to enact the *ObjectiveMissedBehavior* if one exists (otherwise the agent falls into a complete stop).

The *ObjectiveMissedBehavior* is what the agent performs if it misses a location (arrives late). The outcome of the overall action depends on this behavior, for example one could code it to abruptly terminate the agent's endeavor, or to simply skip to the next location and possibly register the failure using some logging facilities. When the last location has been reached, the agent stops after fulfilling all assignments.

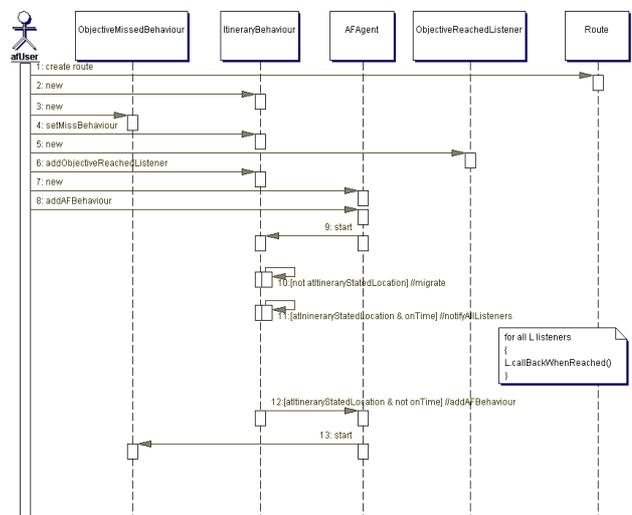

Fig. 4. Itinerary behavior sequence diagram

## III. CASE STUDY: VLE ASSESSMENT SERVICE

Nowadays e-learning virtual environments emerge as useful domains available on the Internet. In exploring the bounds and expressive power of the behavioral patterns presented in the previous section we chose to approach one specific aspect related to VLE, namely the student assessment service (AS). Its design and implementation were employed within the context of an already existing VLE system we had previously created - a web-based virtual university [3].

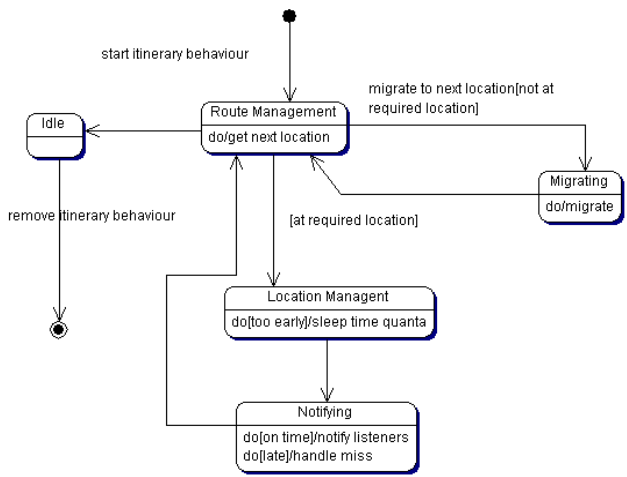

Fig. 5. Itinerary behavior state chart

One of the major problems of creating a virtual environment involves traditional domain modeling and implementing such abstractions using the most suitable technologies [4]. Our approach models a closed organization (no alien/rogue agents are allowed), containing benevolent, cooperative agents that are deployed in a particular configuration and working together to enact the AS.

In what follows we try to prove that employing behavioral patterns is a facile, elegant and fast development agent based solution. Some adjacent design considerations are included as well (i.e. the data source), though we shall not insist on presenting them thoroughly since that would be beyond the scope of our case study and would impair conciseness. Moreover, we want to emphasize the fact that the solution is bound to no specific platform, thus making its porting to another environment a fairly quick and easy task, which implies modifications only at the adapter level (or even just adapter usage, if such adapters already exist for the desired platform). In our case the JADE [7] agent platform has been chosen as supporting mobile agent platform for the assessment service.

The AS provides a means for evaluating a learning entity's (i.e. a student) acquired knowledge and, at the same time, for providing feed-back on her progress. The teaching authority is the one responsible for providing the tests, while the learning entity is the one the tests are addressed to. A test comprises of a number of questions, each having a weight within the total score. To enable automatic evaluation, we use only questions with exact answers, whether they are single choice, multiple choice, true/false, or filling in the blank space (numeric and text are treated separately). A key issue is that the AS is independent from the counterpart system that provides the student with the learning material, thus offering a great deal of flexibility in deploying integrated VLE solutions.

Although tests can have various types, yet they all fall into one of the two approaches considered below:

- a pull scenario, the self-assessment approach, when the learning actor is initiating the assessment to poise her current level of knowledge. The test results are not to be stored within the system, yet for individual progress monitoring, some data is stored.
- a push scenario, i.e. the compulsory exam, initiated by the teaching authority (directly or indirectly) enforcing one particular test to assess all students' knowledge level. The test results shall be stored by the system.

From a multi-tier architectural perspective the agents and their behavioral patterns are part of the business process management layer, while tests and all other persistent objects are stored into a repository and accessed via a database management layer. Presentation components include the GUI and user interaction capabilities agents are capable of.

Embracing a concise tone, in the following few paragraphs we shall confine ourselves to presenting the student oriented part of the AS, leaving aside the teaching authority's interaction with the system. Consequently, the agents' social order and their bearing behaviors for the self-assessment and for the compulsory exam scenarios shall be portrayed next. We believe that these two examples are sufficient for the reader to grasp the features offered by the reusable agent structures.

Throughout the explanations (regarding agents and behaviors) to follow, one should keep in mind that we are dealing with the case of an unbalanced distribution of computational resources due to the locality of the persistent data stores. Hence, a server side is defined as the host where the data is, while a client side is any host where an agent platform has been deployed. Agents are found on both sides, and there are some that migrate from one to another. On both sides we have tried to minimize the number of all-time active agents, managing to reach the point where there are no such objects on the client side. This is a fairly good thing since dormant, all-time active agents on client machines are unacceptable (due to resources overhead and inadequate schedule). Thus, clients activate solely the agent platform and even that only when needed (taking a self-assessment) or required (awaiting for an exam).

For the mandatory tests (i.e. midterms), agents bearing the *Observer* behavior exist on the server side for each exam planned by the teaching authorities (sure enough, one professor can schedule several tests that will ultimately be handled by the same agent, but we shall not approach such details). The pattern allows one such agent to check periodically if the time for the evaluation has come. If so, an *Itinerary* behavior is enacted (by the same agent or by the newly created one) and the agent starts its journey to deliver the tests. This solution is used to release the client platforms from the overhead of dormant agents (see above). Once the agent has reached a location in good time it spawns a new user agent, sets it up with a *Task* behavior that takes over the delivery of the test to the user, detaches and starts its migration towards the next objective. After the last location is visited, the agent returns at the starting point where it awaits test solutions. Data concerning missed/unreachable objectives is gathered as well. The test related material (questions, choices and answers) is packed and carried along, ultimately being copied to each user agent that stays behind to trigger and conduct the assessment. Upon completion of the last action the user agent's *Task* behavior stipulates that it is to send the student's answers back to the server agent that spawned it, and then terminate itself. Once the test solutions arrive from all valid clients and are safely stored in the data repository, the server agent ceases its existence. Note that the client agent is the one that interacts with the user through the GUI and also the one containing the evaluation engine, thus performing the assessment on the spot.

Since the self-assessment scenario requires a user to commence the interaction, on the client side there is no issue regarding permanent dormant agents. It is only the server side that has to start a permanently active agent (with a *Server* behavior deployed on top) that delegates the initial requests from clients (to server workers). It does so by creating a new agent for every requesting entity and handing the responsibility of continuing to serve the client over to them. All subsequent communication between the client and the server is taken care of by the newly spawned agent from this moment on. By contrast, the learning actor initially creates an agent with a *Client* behavior that initiates the communication with the server, revealing the self-assessment intention. Once the communication is established, two virtual channels are created between the two counterparts, one for commands and one for data. To support this, a pair of *Server* and *Listener* behaviors are composed within a *Parallel* behavior on server side, while a *Client* and a *Task* behavior are deployed in the same concurrent manner on the client side. The client-server pair deals with data communication, while the listener-task pair handles the commands (the task is responsible for triggering the listener with a message containing the command). Amongst the valid commands we mention the retrieval of the list of possible valid self-assessment tests, the retrieval of one test, the sending back of the results, and the termination of the session. Actions are performed accordingly employing the necessary data structures. The evaluation engine is integrated within the client and the results are computed on the spot.

In what was previously presented no details concerning how the behaviors are actually deployed were included due to space limitations. One must imagine that the event listeners, triggers and other collaborating parts are coded together throughout each scheme very naturally once the necessary behaviors to endow the agent with are chosen. Hence, the problem faced by the developer has shifted towards the proper orchestration of behavioral patterns, since it is known that computation is orthogonal to coordination [13].

## IV. CONCLUSION AND FUTURE WORK

In this paper we have tried to present how we envision ways of unifying mobile agent platforms through the use of reusable patterns. A new vocabulary of abstractions had to be enforced, one broad enough to encompass the major concepts required by all possible underlying environments. Since one of the initial requirements was that one should be able to change almost everything about agents instantaneously, behaviors are employed to enact agent activity. Our reusable structures are thus behavioral patterns that can be the building blocks of even more complex composite objects. We have identified a number of basic patterns, provided composition mechanisms for these patterns, and created a model that is extensible enough for enabling the developer to add new structures with a minimal amount effort. The framework ensures low coupling between its high abstractions (the patterns) and the sustaining agent platforms through the use of adapters.

Since the starting point of our analysis was a VLE, it is only natural that the strengths and weaknesses of the framework were eventually tested within this context (by means of the described assessment service, to be exact). So far, the model has proved itself to be both fairly flexible and usable, thus adding a significant building block to the path towards different design approaches in the world of mobile agents (though we are aware that using it can be somewhat cumbersome, at least until some "hands on" experience is gained).

We are aware that many improvements are possible and, in the future, we intend to enhance the behavioral patterns and make them easier to use. Also we shall try to provide support for more agent platforms as well. Another important step would be a quality-oriented enhancement of each pattern, with the most stringent aspect being that of better delay computation at the time of agent migration (an issue that appears particularly in the case of the *Itinerary* behavior).

## V. REFERENCES

[1] M. N. Huhns and M. P. Singh, *Agents and multiagent systems: Themes, approaches, and challenges*, chapter 1 in [2], pp. 1-23, 1998.

[2] M. N. Huhns and M. P. Singh, *Readings in Agents,* Morgan Kaufmann, San Francisco, 1998.

[3] T. Marian and B. Dumitriu, "Web based virtual university," Technical University of Cluj-Napoca, 2003. Technical Report.

[4] M. Dinsoreanu, C. Godja, C. Anghel, I. Salomie and T. Coffey, "Mobile Agent Based Solutions for Knowledge Assessment in eLearning Environments", in *Proceedings of the 2003 Euromedia Conference,*

[5] E. Gamma, R. Helm, R. Johnson and J. Vlissides, *Design Patterns,* Addison Wesley Longman Publishing, 1994.

[6] The Foundation of Intelligent Physical Agents (FIPA), http://www.fipa.org/

[7] The Java Agent Development Framework (JADE), http://sharon.cselt.it/projects/jade/

[8] The Agent Development Kit (ADK), http://www.tryllian.com/technology/product1.html

[9] DeLoach S.A. 2000, "Specifying agent Behavior as Concurrent Tasks: Defining the Behavior of Social Agents", *AFIT/EN-TR-00-03*. Technical Report.

[10] F. Zambonelli, N.R. Jennings, A. Omicini, M. Wooldridge, "Agent-Oriented Software Engineering for Internet Applications", in *Coordination of Internet Agents: Models, Technologies and Applications,* 2000, Springer-Verlag.

[11] F. Zambonelli, N.R. Jennings, M. Wooldridge, „Organisational abstractions for the Analysis and Design of Multi-Agent Systems", in P. Ciancarini and M. Wooldridge, editors, *Agent-Oriented Software Engineering,* Springer-Verlag Lecture Notes in AI Volume 1957, January 2001.

[12] Weiss G, *Multi-Agent Systems, A Modern Approach to DAI*, MIT Press, 1999.

[13] D. Gelernter, and N. Carriero, "Coordination Languages and their Significance", *Communication of the ACM,* 35(2), 1992, pp. 97-107.